\documentclass[
final            
  ]
  {aipproc}

\usepackage[letterpaper]{geometry}
\layoutstyle{8x11double}
\usepackage{comment}

\renewcommand{\thetable}{\Roman{table}}

\newcommand{\figRef}[1]{Fig. \ref{#1}}
\newcommand{\tableRef}[1]{Table \ref{#1}}
\newcommand{\insertFigure}[4]{
\begin{figure}[t]
	\centering
  \includegraphics[width=#3]{#1}
  \caption{\label{#2}#4}
\end{figure}
}

\begin{document}
\author{Benjamin Pollard}{
	address={Department of Physics, University of Colorado Boulder, Boulder, CO 80309},
	altaddress={JILA, University of Colorado Boulder, Boulder, CO 80309}
}

\author{Noah Finkelstein}{
	address={Department of Physics, University of Colorado Boulder, Boulder, CO 80309}
}

\author{H. J. Lewandowski}{
	address={Department of Physics, University of Colorado Boulder, Boulder, CO 80309},
	altaddress={JILA, University of Colorado Boulder, Boulder, CO 80309}
}


\title{Studying Expert Practices to Create Learning Goals for Electronics Labs}


\begin{abstract}
Laboratory courses for upper-division undergraduates often involve sophisticated equipment, relatively small class sizes, and extended hands-on projects.
These courses present distinct challenges and opportunities for the physics education research community as these features are not often present in other undergraduate courses.
Here, we focus on an upper-division lab-based electronics course.
As a first step in establishing learning goals for upper-division electronics, we interviewed graduate students and faculty at the University of Colorado Boulder about the use of electronics in their own research labs.
The content-specific nature of electronics courses parallels the hands-on experience of graduate student researchers, so focusing on the experiences of graduate students is ideal for informing lab course reform.
From their interview responses, we developed a framework for classifying applications of electronics.
We identify five types of use and four forms of interaction with electronics content that are consistently identified by faculty and graduate students.
However, we see variations between faculty and graduate students regarding how electronics is learned.

\end{abstract}

\maketitle

\section{Introduction}
\begin{table}
  \begin{tabular}{  p{1.7cm}  p{7.5cm}  p{5cm}  }
  \hline
  \textbf{Type} & \textbf{Definition:} Thinking about or working with... & \textbf{Example} \\ 
  \hline
  Analog & ...a continuous range of voltages, at the single component level. & A servo loop to lock a laser to a cavity \\ 
  Digital & ...a binary voltage signal. & A moving mirror synchronized via TTL pulses \\ 
  Programming & ...a programming language, abstracted away from the underlying voltage levels in the device being programmed. & An automated DC conductivity measurement \\ 
  Black Box & ...a piece of equipment someone else built, beyond the single component level. & 
A spectrum analyzer \\ 
Safety & ...policies, procedures, or devices intended to reduce the risk of bodily harm. & Which precautions apply when changing a laser diode \\ 
\hline
  
  
  \hline
  \textbf{Interaction} & \textbf{Definition:} Work that involves... & \textbf{Example} \\ 
  \hline
  Design & ...planning a strategy to solve a problem by weighing multiple configurations or possible elements and their associated trade-offs. & Designing a rugged voltage probe to use on a rocket \\ 
  Build & ...assembling someone else's design by combining preselected elements. & Hooking up a motor control board \\ 
  Use & ...deploying something that the student did not build himself or herself. & Adjusting servo feedback parameters to control the temperature of a hot plate \\ 
Fix & ...re-visiting design decisions and/or re-building a subset of something in light of a new problem with an existing system. & Isolating and replacing an op-amp in a damaged microscope \\ 
\hline
  \end{tabular}
  
  \caption{\label{definitions}Coding definitions for classifying an instance of electronics use in the lab. An instance can be classified along two dimensions: Type and Interaction. Each dimension has several categories, which are not mutually exclusive.}
\end{table}
Most undergraduate physics programs, including the one at the University of Colorado Boulder (CU), have at least one upper-division lab course. 
These courses present unique learning challenges to students by combining theoretical knowledge with hands-on experience and, in many cases, independent project-oriented work.
To physics education researchers, upper-division lab courses represent an intersection of lab work, upper-division content, and technology-related skills, which are all areas in need of further study \cite{NationalResearchCouncil2012}.
We are beginning the process of transforming the upper-division lab-based electronics course at CU in order to study those areas, as well as to better serve our students.
Our work is informed by studies of the conceptual understanding of electronic circuits, as discussed in ref. \cite{Stetzer2013}. 

Our work follows efforts in transforming advanced labs \cite{SEIreference}, including the senior-level lab course in the CU physics program known as the Advanced Lab \cite{Zwickl2013a}. 
That work established a process for transforming lab-based courses \cite{Zwickl2013}.
The first step in that process was to develop learning goals, which was done by interviewing faculty 
about their views on the existing course, lab courses in general, and the characteristics they looked for in promising students starting in their research labs \cite{Zwickl2012}.

Here, we refine the process of upper-division lab reform, applying it to the junior-level electronics course at CU.
The existing electronics course consists mainly of lab work in small groups, but also involves hour-long lectures twice per week.
The lectures are intended to parallel the lab activities.
Students progress through a series of lab experiments on analog and digital circuits during weekly scheduled lab time and on their own, guided by written lab guides and the instructor.
The course culminates with self-selected projects spanning four weeks.

In transforming the electronics course at CU, we aim to make the course more engaging and inspiring, as well as a better preparation for a research career.
In this sense, we define ``research career'' broadly to include work in academia, industry, government labs, or not-for-profits. 
We begin by focusing on academic applications.
We believe these to be a strong proxy for electronics applications in national labs, industry, and not-for-profit work as many of the techniques are common to all areas.
We conducted interviews of graduate student researchers and research faculty about work in their labs, placing a focus on graduate students' use of electronics.

Three main conclusions emerged from our interviews and subsequent analyses.
First, by focusing on graduate students, we were able to investigate the use of electronics more directly than by focusing on faculty.
Graduate students are a valuable source of information for studying the experiences of experimentalists.
This focus is especially applicable towards electronics since proficiency in electronics depends largely on specific content knowledge.
Second, by analysing interviews of graduate students, we developed a framework for classifying the range of electronics applications present in research labs. 
Applying our framework yielded similar trends among both faculty and graduate students, however we found some key differences between the two groups in perceptions of how electronics is learned.
And third, by analysing our interviews of graduate student researchers, we quantified the relative frequency of various kinds of electronics work. 
By transforming the electronics course to better reflect the distribution of real-world electronics usage, we can better prepare undergraduate students for a research career.

\section{Methods}
To begin developing learning goals for electronics courses, we interviewed graduate students at CU about their use of electronics in their research. 
We solicited a broad range of students covering different sub-disciplines in physics and times spend as graduate students.
We solicited 32 experimental graduate student researchers and interviewed the 22 students who responded.
The respondents were in year 3.4 of their graduate work on average. 
27\% of respondents were female.
8 respondents worked in atomic, molecular, and optical (AMO) physics, 7.5 in condensed matter, 3.5 in space/plasma, 2 in high energy, and 1 in biophysics, with half counts accounting for work spanning two fields.

The interviews were conducted in a semi-structured format.
The bulk of the interview was comprised of responses to the question, ``What is the most important aspect of electronics that you need to know for your current research?''
Interviewees ultimately spoke about several instances of electronics use.
For each instance, we heard about the type of electronics used, the way in which the student interacted with electronics, and in what context the student learned the skills needed to carry out the work.
We asked directed questions to fill in the gaps once the interviewee had volunteered information for each instance.

We took audio recordings and written notes during the interviews, which were analysed afterwards.
From this analysis, we developed and evolved categories for classifying students' responses that we used to code the interviews.
After the definitions were finalized, records from 20\% of student interviews were re-coded by an additional researcher in PER who was not involved in this work, as a test of inter-rater reliability.
There was a 90\% agreement in coding before discussion, and a 99\% agreement afterwards.

Subsequent to the graduate student interviews, we conducted interviews of nine physics research faculty members at CU (of 12 faculty solicited).
These faculty were comprised of one assistant, three associate, and five full professors.
All of these interviewees had been instructors of the CU electronics course at some point within the last four years, and had experience in experimental research labs spanning fields of AMO, condensed matter, and high energy.
We asked about the use of electronics by graduate students in their research lab using a format very similar to the graduate student interviews.
Note that only two of our student interviewees worked in the lab of one of our faculty interviewees.
Nonetheless, our representative sample of graduate students who use electronics still makes for an illuminating comparison with our sample of undergraduate electronics instructors as the research fields of the two sample groups are similar.

\section{Results and Discussion}
Analysis of our interviews resulted in a framework for classifying the use of electronics in a research environment.
\tableRef{definitions} shows our categories of electronics use along two distinct dimensions, with accompanying definitions and exemplary instances.
The two dimensions we used to classify electronics work were \textit{Type}, referring to the type of electronics content knowledge needed for the work, and \textit{Interaction}, referring to how researchers interacted with their equipment. 
Of the five categories along the Type dimension, the first four were identified before coding, while Safety was added during coding as an emergent category.
Similarly, Fix was added as an emergent category in the Interaction dimension.
Note that while each category is distinct from the others, they are not meant to be mutually exclusive.

Among the Interaction categories of Design, Build, Use, and Fix, the first three form a ordered progression. 
In our coding, it was clear that Designing was always soon followed by Building, and likewise Building was followed directly by Using.
Therefore we decided to categorize each instance as only the earliest category in this progression.
Our scheme forces counts in the Build and Use categories to represent only instances in which the researcher picked up the progression sometime after the Design phase.
We believe that Design understanding is generally broader and more complete than the understanding required to Build a design, while the understanding required to Use is narrower still.
The understanding required to Fix, however, is harder to rank among the other categories, but often reaches a level of understanding akin to Designing, albeit from a different angle.

Some combinations of Type and Interaction categories are incompatible with each other, and thus were explicitly forbidden in our coding.
Designing and Building does not make sense for a Black Box, given our definition of a Black Box.
While Designing and Fixing a Program are conceptually distinct, in practice it is impossible to distinguish between those two Interactions, since the programmer constantly flips between Designing and Fixing during his or her work.
Therefore, we decided to code all of those cases as Programming Design, effectively excluding Programming Fix, to avoid ambiguity.
While Programming Building was not explicitly forbidden, there were no cases in which a graduate student researcher received a Programming Design from someone else and then Built it him or herself.

\insertFigure{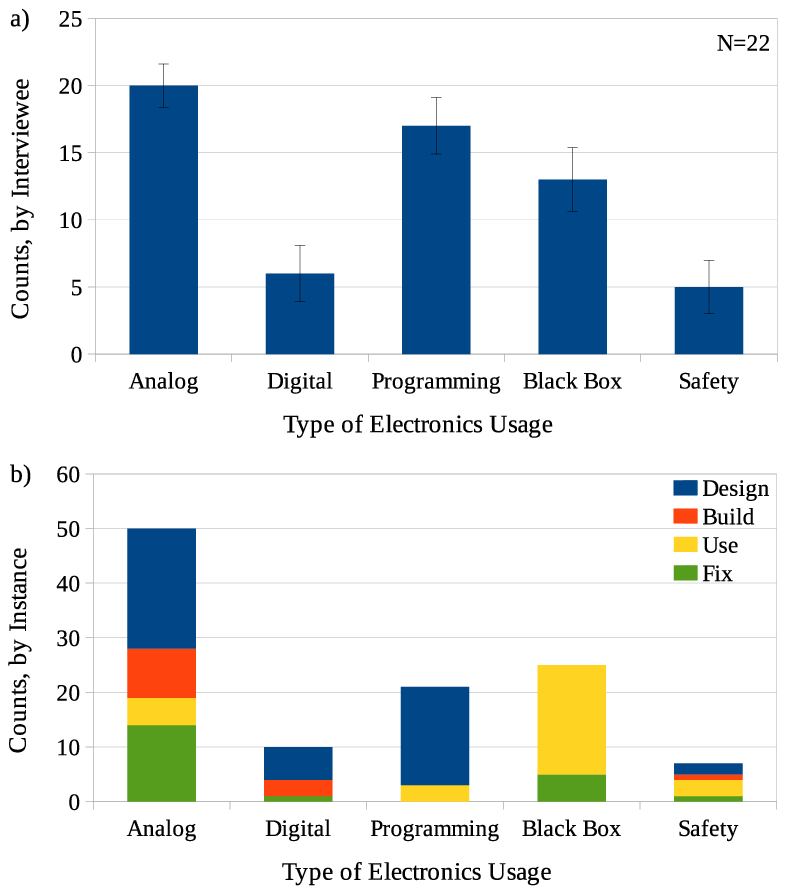}{comparisons}{\columnwidth}{Graduate students' work with electronics, counted by interviewee (a) or by instance (b) and categorized by Type of use. Colors in (b) indicate the breakdown of each type by Interaction. Error bars represent the standard error of the sample proportion.}
The frequency of graduate students who mentioned work in each Type category is shown in \figRef{comparisons}a. 
These results suggest that analog electronics is the most prevalent type of electronics used by physics researchers, while programming and the use of black boxes (such as commercial test equipment) are also common.

Comparisons and relationships between dimensions can provide insight into the best approaches for teaching and learning electronics.
For example, \figRef{comparisons}b shows the Type of electronics used by graduate students colored by their Interaction with that piece of electronics.
We constructed a contingency table from this information and conducted Pearson's Chi Squared Test of Independence \cite{agresti2009statistical} to see if these two dimensions are statistically independent.
We eliminated the Programming and Black Box categories from this analysis because those categories were affected by category combinations explicitly forbidden in our coding scheme. 
We found a p-value of 61\%, indicating little relationship between the Type of electronics and researchers' Interaction with it.
This suggests that student researchers Interact with electronics in a variety of ways over the course of their work.

\insertFigure{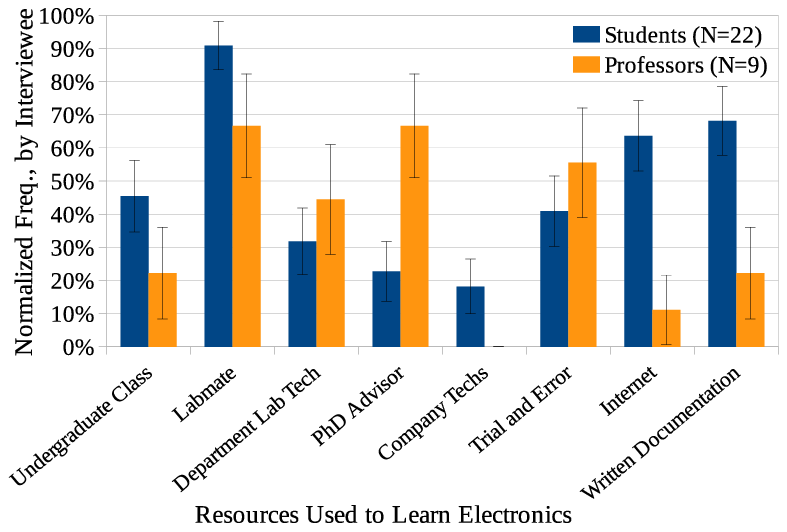}{howLearned}{\columnwidth}{The frequency of resources graduate students used to learn electronics skills. Students and professors were asked how graduate students learn the skills they use in lab. The number of people who fell into each category are shown, normalized by the total number of people interviewed. Error bars represent the standard error of the sample proportion.}
In addition to classifying what and how electronics is used in research labs, we asked our interviewees how graduate students learn the skills they need for their electronics work.
Perceptions of how students learn electronics are shown in \figRef{howLearned}. 
Our analysis shows a comparison of graduate student and faculty perceptions.
Eight categories emerged from the responses, as shown along the horizontal axis of \figRef{howLearned}.
We counted the number of interviewees who mentioned each category and respectively normalized those counts by the total number of graduate students or faculty interviewed.
The perceptions of students and faculty roughly match with a few notable exceptions.
More faculty than students responded that students learn from their advisors, while more students responded that they learn from the Internet, company technicians, and the written documentation pertaining to their lab equipment.
These differences illustrate the benefit of focusing on graduate student experiences when investigating common practice in research labs.

\section{Conclusions}
By interviewing graduate students and faculty on their use of electronics in their research labs, we gained a sense of the use of electronics in modern physics research.
These findings can form the basis for learning goals and the overall transformation process for our upper-division lab-based electronics course.
Analysis of our interviews led to a classification scheme for the use of electronics in the lab.
The frequency of each category paints a picture of what electronics lab work looks like, which can be used to better match an undergraduate course to professional research.
For example, Analog electronics is the most prevalent Type of electronics used by physics researchers, while Programming and the use of Black Boxes (such as commercial test equipment) are also common.
Furthermore, no single Interaction was associated with Analog, Digital, or Safety electronics. 
These findings suggest that a variety of Types and Interactions should be well represented in electronics curricula.

Graduate students were an important demographic to study in learning about the use of electronics in research, as it is these people who are most familiar with the specific content and skills needed in the lab.
Content-specific courses such as upper-division electronics could benefit greatly from perspectives gained through graduate student researchers.
While graduate students spoke of the same categories of electronics as faculty, some discrepancies existed in perceptions of how electronics is learned. 
These discrepancies may indicate, for example, that electronics instructors should make a conscious effort to encourage the use of the Internet, written documentation, and experts outside the classroom (as a proxy for company technicians) to learn electronics.

The classification scheme developed here could be modified and extended to other areas of lab work besides electronics, for example optics, to achieve the same goal of better preparation for professional research.
We intend to use the findings of this work as a starting point for developing learning goals for a lab-based electronics course.
In parallel, we will broaden and refine our investigation of the usage of electronics in research labs by developing a survey suited for dissemination to other institutions beside CU.
Responses from this broader base of respondents will paint a clearer picture of the role electronics plays in the varied landscape of modern physics labs.


\section{Acknowledgements}
We thank Bethany Wilcox for coding for inter-rater reliability. We also thank the CU graduate students and faculty members who volunteered their time for interviews. This work is supported by NSF DUE-1323101.

\bibliographystyle{aipproc_bp}
\bibliography{references}

\end{document}